\documentclass[final,5p,times,twocolumn,numbers]{elsarticle}

\usepackage[numbers]{natbib}
\usepackage[superscript]{cite}
\usepackage[hyphens]{url}
\usepackage{graphicx}

\journal{...}

\begin{document}

\begin{frontmatter}




\title{Cutting Through the Confusion and Hype: Understanding the True Potential of Generative AI}

\author{Ante Prodan\textsuperscript{a,b,c,d,*}}
\cortext[cor1]{Corresponding author. Email: a.prodan@westernsydney.edu.au}
\affiliation{School of Computer, Data and Mathematical Sciences, Western Sydney University, Sydney, Australia}
\affiliation[first]{Brain and Mind Centre, University of Sydney, Camperdown, Australia}
\affiliation{Mental Wealth Initiative, University of Sydney, Camperdown, Australia}
\affiliation{Computer Simulation & Advanced Research Technologies (CSART), Sydney, Australia}

\author{Jo-An Occhipinti\textsuperscript{b,c,d}}

\author{Rehez Ahlip\textsuperscript{a}}

\author{Goran Ujdur\textsuperscript{b,c}}

\author{Harris A. Eyre\textsuperscript{c,e,f,g}}
\affiliation{Brain Capital Alliance, San Francisco, California, USA}
\affiliation{Baker Institute for Public Policy, Rice University, Houston, Texas, USA}
\affiliation{Meadows Mental Health Policy Institute, Dallas, Texas, USA}

\author{Kyle Goosen\textsuperscript{d}}


\author{Luke Penza \textsuperscript{a}}

\author{Mark Heffernan\textsuperscript{a,d,f}}
\affiliation{Dynamic Operations, Sydney, Australia}





\begin{abstract}
This paper explores the nuanced landscape of generative AI (genAI), particularly focusing on neural network-based models like Large Language Models (LLMs). While genAI garners both optimistic enthusiasm and skeptical criticism, this work seeks to provide a balanced examination of its capabilities, limitations, and the profound impact it may have on societal functions and personal interactions. The first section demystifies language-based genAI through detailed discussions on how LLMs learn, their computational needs, distinguishing features from supporting technologies, and the inherent limitations in their accuracy and reliability. Real-world examples illustrate the practical applications and implications of these technologies.

The latter part of the paper adopts a systems perspective, evaluating how the integration of LLMs with existing technologies can enhance productivity and address emerging concerns. It highlights the need for significant investment to understand the implications of recent advancements, advocating for a well-informed dialogue to ethically and responsibly integrate genAI into diverse sectors. The paper concludes with prospective developments and recommendations, emphasizing a forward-looking approach to harnessing genAI's potential while mitigating its risks.

\end{abstract}

\begin{keyword}
artificial intelligence AI, large language models, AI and society 
\end{keyword}

\end{frontmatter}




\label{introduction}
\begin{center}
\section*{“If economics wants to understand the new economy, it not only has to understand increasing returns and the dynamics of instability. It also has to look at cognition itself, something we have never done before in economics.”}
\end{center}
\begin{flushright}
— W. Brian Arthur, "Coming from Your Inner Self". www.presencing.org. April 16, 1999. 
\end{flushright}
Artificial Intelligence (AI) and specifically generative AI (genAI) occupies a unique space in the public consciousness, often shrouded in a mix of gross underestimation of its long-term impact, breathless short-term hype, dystopian fears, and genuine misunderstanding. GenAI undoubtedly represents a transformative force in technology, but its capabilities and limitations are frequently misconstrued.    

As a consequence, policymakers, businesses, and the public may either overstate the immediate effects, leading to undue fear and potentially stifling innovation, or underestimate the long-term implications, resulting in a lack of preparedness for the systemic changes it will bring about. In either case, such misconceptions hinder the development of effective strategies for integrating genAI into society in a way that maximises its benefits while mitigating its risks. It is therefore imperative that the dialogue surrounding genAI be grounded in a balanced understanding, informed by empirical research and thoughtful analysis, to navigate its integration responsibly and ethically.

While the ability to generate realistic images, music and videos has captured widespread attention, our focus lies in the domain of language-based genAI, particularly chatbots. Why this emphasis on language? While visual generative AI excels in rapid visual content creation, language serves as the foundation for communication, information exchange, and knowledge work. The ability to generate human-quality text opens doors to a far broader range of applications, from revolutionising customer service and automating business processes to personalising education and enhancing knowledge work and human creativity. 

This paper is designed for a broad readership. The first section delves into the exciting landscape of language-based genAI, exploring its core concepts and supporting technologies. Through a series of fact-checks, we aim to demystify genAI, allowing readers to evaluate common but frequently inaccurate statements while dispelling deeper misconceptions and unfounded claims. By doing this we address the following topics:
\begin{itemize}
  \renewcommand{\labelitemi}{$\rightarrow$} 
  \item Understanding how Large Language Models (LLMs) learn.
  \item Comprehending the scale of data and computation used.
  \item Distinguishing between LLMs and supporting technologies that are used to construct a chatbot.
  \item Grasping the intrinsic limitations in accuracy and reliability.
  \item Recognising the role of prompt engineering.
\end{itemize}

Whenever possible, we provide real-world examples to practically illustrate the impact of the introduced concepts and the profound influence genAI is poised to have on our interactions with machines and each other.

In the second section, we focus on a systems perspective, examining concrete scenarios of combining and recombining   LLMs with existing technologies and their effect on productivity. Finally, we highlight several key areas of concern that underscore the need for increased investment in understanding the implications of recent advancements in genAI. We conclude by offering a succinct overview of possible future developments and principle recommendations. We hope this paper provides clear, informed, and nuanced insights into this powerful technology, fostering a better understanding of the diversity of issues associated with generative AI.

A note on terms used and typography: When referring to concepts like 'intelligence,' 'reasoning,' or 'understanding' in the context of genAI, we do not attribute human-like meanings to these terms. Instead, we use them to describe AI capabilities that mimic human abilities in completing specific tasks.  We acknowledge that this approach may lack precision, but we favour it over using complex technical jargon which may be inaccessible to a general readership. We employ bold font  to emphasise key concepts and statements, particularly those whose implications are often misunderstood.

\label{Fact checking }
\section*{Fact checking }
\subsection*{Large Language Model (LLM)}
An LLM is a complex AI system trained on very large amounts of text data (Figure 1 step 3), of a trillion or more tokens , including books, articles, computer code, and web pages. To illustrate the magnitude of this data, a trillion tokens would equate to a content of   approximately 4 million books, similarly the compute required for LLM’s inference on the input of 1000 tokens is equivalent of compute required for common usage of Microsoft Excel for years. Finally, the compute required for training of LLMs requires an amount of electricity that is sufficient to run a small town over weeks or months. This extensive training enables an LLM to understand, generate, and manipulate human language with remarkable proficiency \cite{improving}. An LLM leverages deep learning techniques, particularly a specific type of deep artificial neural network (DNN) called transformers \cite{attention}, to learn the intricacies of language. After the training an LLM can be used, for example, as part of a chatbot to respond to user inputs as well as to perform various language-related tasks such as content creation, summarisation, or translation. All these tasks generate tokens, hence the name GenAI. In the period from 2017 the development of genAI has accelerated, leading to human level performance in different areas \cite{progress}.  

\begin{figure}[ht]
\centering
\includegraphics[width=0.48\textwidth]{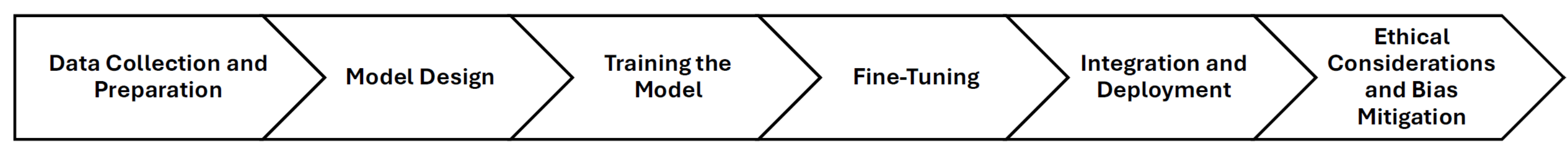}
\caption{LLM development cycle. Each of the steps requires careful planning, a team of experts in machine learning, data engineering, and domain-specific knowledge to successfully develop and deploy a model like GPT-3. Training requires large computational resources.}
\label{fig:image1}
\end{figure}

\subsection*{Artificial Neural Networks (ANNs)}
ANNs are computational models inspired by the structure and function of the human brain. They consist of interconnected nodes, also known as artificial neurons, organised in layers. While they are based on some baseline similarities, artificial neurons are different in details of structure and function to human or animal neurons. Each neuron receives input signals, processes them, and transmits output signals to other neurons. The connections between neurons have associated weights that determine the strength of the signal transmission, these numbers are also known as parameters of an LLM. Through a process of training on data, these weights are adjusted, allowing the network to learn and recognise patterns.

\begin{figure}[ht]
\centering
\includegraphics[width=0.48\textwidth]{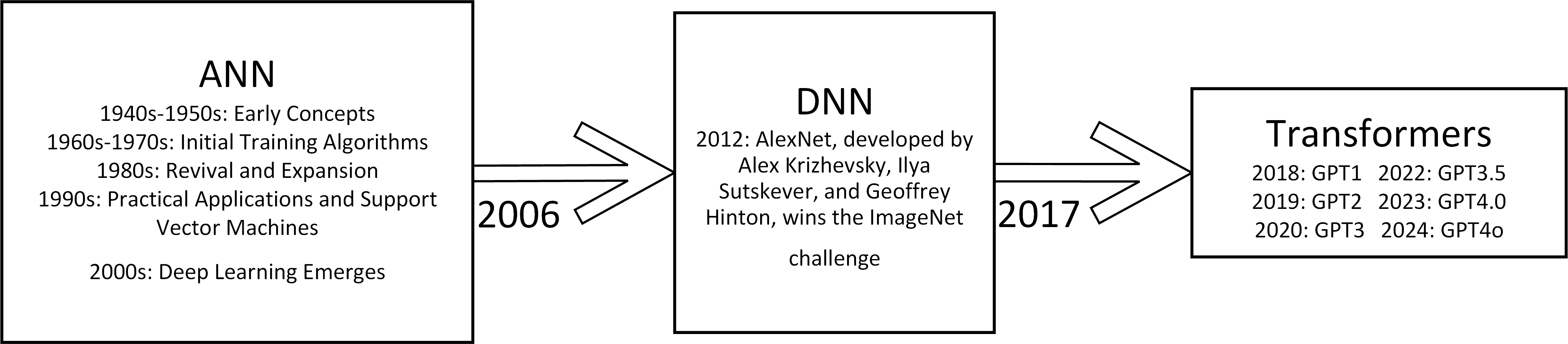}
\caption{Evolution of Artificial Neural Networks that enabled creation of LLMs}
\label{fig:image2}
\end{figure}

\subsection*{Deep neural networks (DNNs)}
Deep neural networks, a subset of ANNs, feature multiple hidden layers of neurons between the input and output layers, which allow them to model complex functions and interactions within the data \cite{deep}. This depth enables the network to perform sophisticated tasks like image and speech recognition, and natural language understanding, much more effectively than shallower networks. The ability to learn   from vast amounts of data resulting in generational improvements over time makes DNNs powerful tools for a wide range of applications, from autonomous driving to medical diagnosis. 
DNNs are often described as "black boxes" due to the difficulty in interpreting their reasoning, which arises from their complex and opaque structures. They incorporate potentially billions of parameters that subtly influence the decision-making process in non-intuitive ways, making it challenging to trace how individual contributions lead to final outcomes. Additionally, DNNs operate through layers that interact in non-linear ways; minor changes in input or parameters can result in significant and unpredictable output variations. This complexity is compounded by hidden layers that transform data in ways not directly accessible or interpretable, further obscuring the understanding of how DNNs reach their conclusions.
As the size of an DNN increases (scales up), it generally becomes more powerful but requires more computational resources. Effective scaling involves the use of techniques like distributed computing and specialised hardware (e.g., Graphical Processing Units - GPUs  ) to manage the increased computational load. Moreover, strategies like pruning (removing redundant neurons) and quantisation (reducing the precision of the calculations) are employed to optimise the network’s efficiency without significant loss of accuracy \cite{scaling}.

\begin{figure*}[htb]
\centering
\includegraphics[width=1\textwidth]{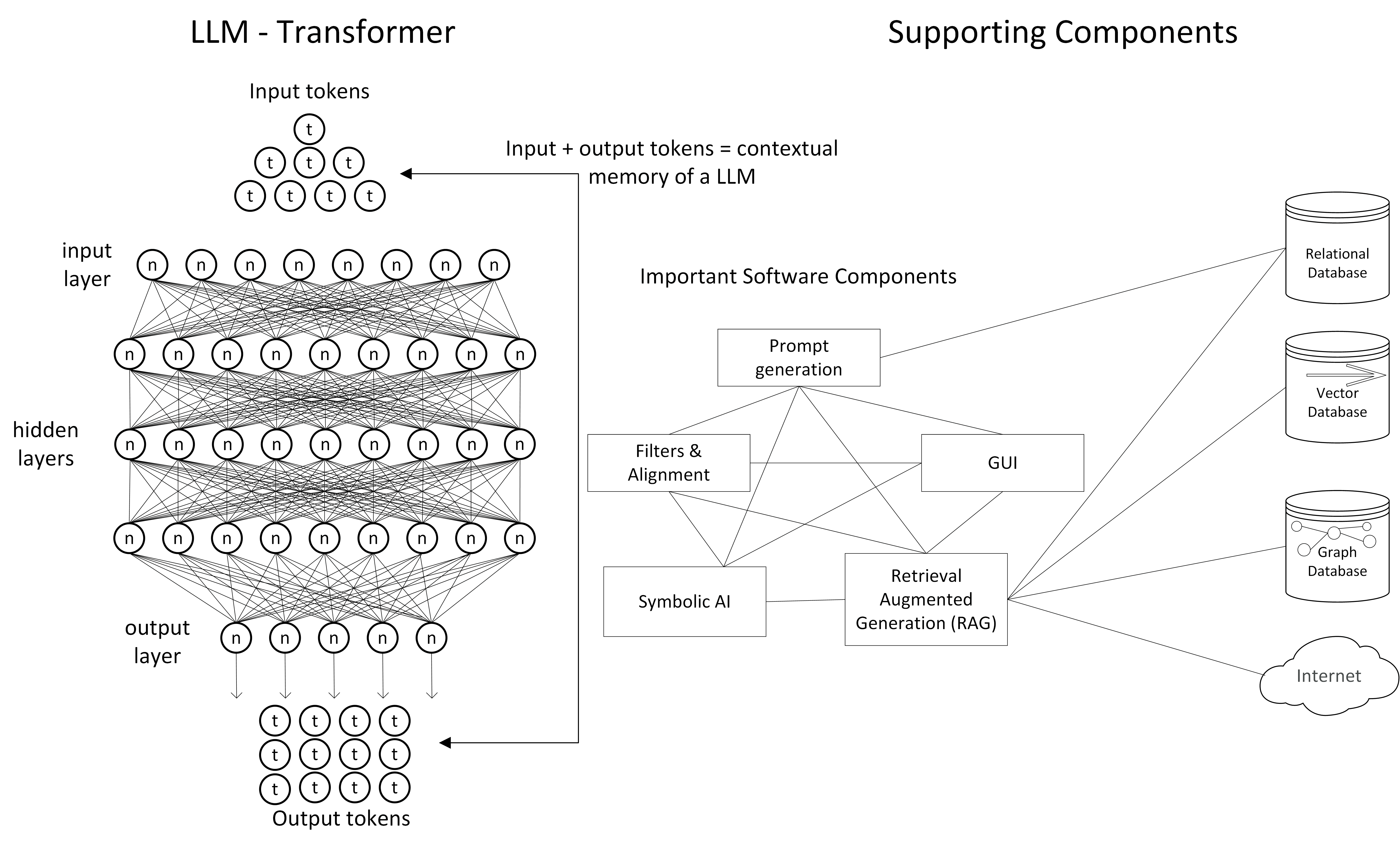}
\caption{Architecture of an AI application (e.g. chatbot).}
\label{fig:image3}
\end{figure*}

\subsection*{Transformers: A Powerful DNN Architecture for LLMs}
Transformers are a type of DNN specifically designed for sequence processing tasks like language understanding and generation. Their key innovation is the self-attention mechanism \cite{attention}, which allows the model to analyse the relationships between all words in a sentence simultaneously. This enables transformers to capture long-range dependencies and context, leading to a deeper understanding of language compared to traditional   DNNs such as recurrent neural networks (RNNs). 
\textbf{An LLM does not learn through use (token generation) and it is not a database.}
As mentioned, LLMs are trained on vast amounts of text data, they don’t store the data itself. Instead, the ‘knowledge’ is encoded within the parameters of the DNN. This process is like a learning process of a human or an animal. Once training is complete, an LLM can respond to prompts and generate text by predicting the most likely sequence of tokens based on the patterns it has learned. The generated text is not retrieved from storage like with a database but generated and therefore not an exact reproduction of the training data.  An LLM utilises information provided in prompts provided by a user, allowing it to generate responses that incorporate new facts or instructions.  
The context size is one of the key operational parameters of a LLM, it is measured in tokens, and all prompting and token generation must be done within this size – an LLM does not have any other form of memory, therefore an LLM cannot permanently learn new information without further training of the original model. Fine-tuning is a process that involves training the LLM on additional data to adapt its knowledge and skills to a specific task or domain, effectively updating the weights of the ANN. This process requires much more data and compute than use by prompting, however this is still multiple orders of magnitude less than what is required for the initial training process. It is important to note that after the fine-tuning, the performance of the fine-tuned LLM in more generalised use can degrade.

\subsection*{LLM Challenges}
LLMs face many challenges, such as biases present in training data leading to biases in generated tokens, hallucinations \cite{hallucinations}, inability to directly apply logic, arithmetic or temporal reasoning, the potential for generating harmful content, limited context size, and very large data, computational and memory resources required for their training, fine-tuning, and deployment \cite{failures}. Some of these challenges are addressed by changes in training of LLMs, through the curation of training data for example, or changes to the underlying ANN by quantisation. Many other optimisation techniques are implemented through separate software components that are integrated with LLMs within chatbots and primarily focus on prompt engineering. 

\subsection*{What is a chatbot?}
A chatbot is a complex software application (see Figure 3) designed to simulate conversation with human users, typically through text-based interfaces. They are used in various applications, including customer service, virtual assistants, entertainment, and education. Chatbots are sophisticated systems, and while LLMs play a crucial role, they are just one piece of the architecture.

\subsection*{The LLM's Role in chat-bot:}
LLMs are the "language brain" of modern chatbots, providing the ability to understand and generate human-like conversation. They enable the chatbot to:
\begin{itemize}
    \item Comprehend user queries: Deciphering the meaning and intent behind user message.
    \item Formulate relevant responses: Crafting natural and engaging responses that address the user's needs.
\end{itemize}

\textbf{In addition to LLMs, chatbots can contain many supporting components that address limitations of LLMs or provide a specific service such as a user interface.}

\label{Essential Supporting Components by function}
\section*{Essential Supporting Components by function}
\begin{itemize}
\item \textbf{Maintain context:} LLMs on their own cannot inherently maintain context across multiple exchanges in a conversation. The chatbot must include components that store the conversation history, including both user inputs and the chatbot's responses allowing the appearance that a bot is remembering previous interactions and providing a more coherent and personalised conversation flow. This highlights the importance of the process called prompt engineering.
\item \textbf{Filter:} Essential for ensuring the chatbot produces appropriate and safe content. Filters can detect and block offensive language, prevent biases and the disclosure of sensitive information, and ensure responses align with ethical guidelines.
\item \textbf{Symbolic AI:} More recent development integrates Symbolic AI and LLMs to creates more robust chatbots, leveraging Symbolic AI's strengths in explicit, rule-based reasoning while using LLMs for conversational fluency and ambiguity management.
\item \textbf{Python code execution:} The Python code generated by LLM is used to solve a specific problem such as data extraction or statistical analysis that is then integrated into the chatbot's response.

\item \textbf{Grounding in Data – Data Retrieval and Integration:}
\begin{itemize}
\item \textbf{Internet Access:} Since LLMs are static, to keep the information provided to the users up to date many chatbots are connected to the internet, allowing them to retrieve up-to date information, news, or other relevant data to enrich their responses.
\item \textbf{Database management \& embeddings and indexation:} Chatbots often interact with databases to access user profiles, previous  conversations, product information, or historical data, enabling them to personalise interactions and provide specific services. To quickly find relevant information within large datasets, chatbots often employ text indexation and search mechanisms including, vector and/or graph databases. This allows them to efficiently access relevant knowledge when responding to user queries. The embeddings have a central role in this process. They are a way to translate words and phrases into numerical codes (vectors) that computers can understand, capturing their meanings and relationships. Embeddings are dense vector representations of text data. They capture the semantic meaning of words, phrases, or even entire documents in a continuous vector space. Each word or phrase is represented as a point in this high-dimensional space, where similar meanings are positioned closer together, and dissimilar meanings are farther apart. Think of it as a digital map where similar concepts are placed close together, making it easier for AI systems to recognise patterns and context. For instance, in a chatbot, embeddings help the system understand the intent behind customer queries and respond appropriately, ensuring more accurate and meaningful interactions. This technology underpins many advanced AI applications, enabling them to process and make sense of large library of documents or other data and use human language efficiently.
\end{itemize}

\item \textbf{Interacting with other systems: Function calling \& Agents:} Some chatbot frameworks allow LLMs to directly call external functions or Application Programming Interfaces (APIs).   This empowers the chatbot to perform actions beyond text generation, such as running any software application with inputs it specifies, booking appointments, making reservations, or controlling smart home devices, making the interaction more practically useful and dynamic. Similarly, agents refer to software components that actively interact with the environment or other systems based on instructions or data processed by the LLM. These agents can be specialised LLMs, and execute actions, make decisions, or managing interactions autonomously, based on the output from the main LLM that is in supervisory role that involves decomposing complex problems and dispatching tasks to agents. In summary, agents are designed for more complex and autonomous interactions with various systems, adapting and potentially learning from their environment. In contrast, function calling is about executing specific tasks directed by the LLM’s output, without the autonomy or complexity involved in managing ongoing interactions or maintaining a conversational state. Both technologies serve distinct roles depending on the required complexity and autonomy of the task at hand and play a critical role in enhancing the functionality and applicability of chatbots beyond mere text generation, bridging the gap between LLM’s language capabilities and practical, real-world applications.
\item \textbf{User Interface:} The chatbot's user interface (UI) can be text-based (like in messaging apps), voice-based, or even graphical. The UI determines how users interact with the chatbot.

\end{itemize}

\label{The Importance of a Systemic View on genAI}
\section*{The Importance of a Systemic View on genAI}
Discussions about genAI often focus narrowly on the capabilities of LLMs, overlooking the essential roles of other technologies such as symbolic AI, reinforcement learning, and multimodal systems. This limited perspective fails to capture the crucial interplay between LLMs and these existing techniques, thereby hindering a comprehensive understanding of genAI's broader impact on work and society. 

To fully grasp genAI's transformative potential, we must adopt a systemic perspective that recognises LLMs as operating within a complex ecosystem of interconnected components. The challenge for genAI is not in its capabilities but in its integration and adoption. The accelerating performance trajectory of genAI derives not only from scaling LLM capabilities but also from the seamless integration and recombination with new and existing software components \cite{arthurOnTech}, ranging from databases to symbolic AI systems. Recent advances in multimodal research across diverse fields—including mathematics, biology, genomics, physical sciences, and neuroscience—provide evidence for the efficacy of this integrative approach.\cite{stateOfAi} A systemic perspective illuminates how genAI will reshape various facets of work and society through non-linear, compounding advancements as we better integrate these advanced technologies into our digital environments.

When evaluating genAI's potential, we must consider three critical dimensions that indicate no evident barriers to rapid development:

\begin{enumerate}
    \item \textbf{Compute Requirements:} 
    Optimisation techniques like quantisation and pruning are significantly reducing operational resource requirements for complex inference at constant model sizes \cite{deepCompression,reducingMemory}, making advanced genAI applications more economically viable for a broader range of entities within the next three years\cite{4charts}. On the other hand, the compute demands for training state-of-the-art LLMs continue to increase exponentially. This dichotomy is highlighted by the following facts. Companies like OpenAI have managed to achieve substantial reductions in compute required for inference— in May 2024, Kevin Scott, Microsoft's CTO, stated that GPT-4 \cite{gpt4technicalreport} achieved a 12-fold decrease in compute required for its use while doubling token generation speed compared to its predecessors. Conversely, the escalating computational resources needed for training large foundational models are becoming exponentially costly, potentially limiting this capability to a few major players such as Google, Meta, and Microsoft. Therefore, rather than focusing solely on current limitations, we need to anticipate genAI's trajectory, acknowledging both the democratisation of inference capabilities and the centralisation of training resources among a few dominant entities.
    
    \item \textbf{Software Integration:} As supporting software components continue evolving, new capabilities will be rapidly integrated into genAI systems. Recent breakthroughs, such as improved mathematical reasoning and the introduction of agents, exemplify this accelerating convergence \cite{stateOfAi}.
    
    \item \textbf{Availability and Development Cycles:} While the initial development cycle for new genAI-based applications can take up to 36 months, this time frame will likely shrink to 12-18 months as the technology matures. Moreover, genAI's self-perpetuating nature, particularly in code generation, holds the potential for much shorter development cycles.
\end{enumerate}

Furthermore, genAI's implementation varies across industries, shaped by technological literacy and economic incentives. Finance and high-tech, with strong digital infrastructures and expertise, are rapidly adopting genAI for personalised experiences and gaining a competitive edge \cite{mckinsey}. Consequently, a systems science-based, forward-looking approach is crucial to fully grasp genAI's potential impact on businesses, governments, and international bodies. Relying solely on historically-based modelling is inadequate due to the technology's non-linear nature and ability to rapidly recombine with existing technologies \cite{arthurOnTech}. Substantial investment is urgently needed to better understand genAI's ramifications across various domains, particularly its potential to disrupt or create human occupations by altering their form, function, and meaning \cite{arthurOnTech, socialImplications}.

As we integrate genAI more deeply into business and societal functions, ethical implications become increasingly complex and critical \cite{bostrom2014ethics}. Our previous work provided a framework for considering the broader societal transition from the Age of Information to the Age of Intelligence \cite{navigating, RN31}. One of the foremost ethical concerns is the risk of exacerbating existing inequalities. GenAI can significantly enhance productivity and economic output, but without careful consideration, these benefits may disproportionately favour the already advantaged \cite{invisible}. Ethically, it is essential to manage AI deployment to prevent increased social stratification and ensure inclusivity. Additionally, improving the transparency of AI decision-making processes is crucial to prevent unintended consequences of opaque algorithmic functioning, fostering greater trust and acceptance.

The rapid development witnessed recently underscores the necessity of proactively addressing potential worst-case scenarios. While difficult to predict precisely, some concerning possibilities include:

\begin{itemize}
    \item Widespread job displacement and economic disruption due to automation outpacing the ability to retrain and re-skill workers \cite{raceMM,recessionarypressures}.
    \item Proliferation of misinformation, fake content, and deepfakes at an unprecedented scale, eroding public trust and social cohesion \cite{deepfakes}.
    \item Existential risks arising from advanced AI systems pursuing misaligned goals or exhibiting unintended behaviours beyond human control \cite{alignment}.
    \item Concentration of power and influence in the hands of a few tech giants or nations, exacerbating inequalities and geopolitical tensions, and undermining democracy \cite{machineAge}. 
\end{itemize}

Mitigating these risks requires a multi-stakeholder approach involving policymakers, technologists, ethicists, researchers, and the broader public. Responsible development frameworks, robust governance structures, and proactive planning for socioeconomic transitions \cite{mentalwealth} will be essential to harness genAI's benefits while minimising potential downsides.

\section*{}
{\itshape This paper solely reflects the views, opinions, arguments of its authors and does not necessarily represent the perspectives of the organisations that authors are associate with.}  

\subsection*{Statement of potential competing interests:}
Authors AP, RA, GU, KG, SS, LP, and MH declare they have no conflicts of interest relevant to this work. Author JO is both Head of Systems Modelling, Simulation \& Data Science, and Co-Director of the Mental Wealth Initiative at the University of Sydney's Brain and Mind Centre. She is also Managing Director of Computer Simulation \& Advanced Research Technologies (CSART) and acts as Advisor to the Brain Capital Alliance. 

\subsection*{Author contribution:}
Manuscript concept and drafting: AP; critical revision of manuscript and contribution of important intellectual content: all authors.

\appendix

\bibliographystyle{unsrt} 
\bibliography{AI_paper_references2.bib}







\end{document}